\documentclass{ws-procs961x669}            

\usepackage{graphicx}
\usepackage{microtype}
\usepackage{amsmath}
\usepackage{amssymb}
\usepackage{subfigure}
\usepackage{hyperref}
\usepackage{makecell}
\usepackage{url}
\usepackage{xcolor}
\usepackage{color}
\usepackage{mathrsfs}
\usepackage{calrsfs}
\usepackage{amsfonts}
\usepackage{lipsum}
\usepackage{tabularx}
\usepackage{eucal}
\usepackage{latexsym}
\usepackage{ragged2e}
\usepackage{epsfig}
\usepackage{textcomp}

\begin{document}

\title{Wormhole geometries in $f\left(R,T^2\right)$ gravity satisfying the energy conditions}

\author{Nailya Ganiyeva}
\address{Instituto de Astrof\'{i}sica e Ci\^{e}ncias do Espaço, Faculdade de Ci\^{e}ncias da Universidade de Lisboa,
Edif\'{i}cio C8, Campo Grande, P-1749-016 Lisbon, Portugal \\ $^*$E-mail: fc57452@alunos.fc.ul.pt}

\author{Jo\~{a}o Lu\'{i}s Rosa}
\address{University of Gda\'{n}sk, Jana Ba\.{z}y\'{n}skiego 8, 80-309 Gda\'{n}sk, Poland \\ Institute of Physics, University of Tartu, W. Ostwaldi 1, 50411 Tartu, Estonia \\ $^*$E-mail: joaoluis92@gmail.com}

\author{Francisco S. N. Lobo}
\address{Instituto de Astrof\'{i}sica e Ci\^{e}ncias do Espaço, Faculdade de Ci\^{e}ncias da Universidade de Lisboa,
Edif\'{i}cio C8, Campo Grande, P-1749-016 Lisbon, Portugal \\ Departamento de F\'{i}sica, Faculdade de Ci\^{e}ncias da Universidade de Lisboa, Edif\'{i}cio C8, Campo Grande, P-1749-016 Lisbon, Portugal \\ $^*$E-mail: fslobo@fc.ul.pt}

\begin{abstract}
We explore the properties of traversable wormhole spacetimes within the framework of energy-momentum squared gravity, also known as $f(R,T^2)$ gravity, where $R$ represents the Ricci scalar, $T_{ab}$ is the energy-momentum tensor, and $T^2 = T_{ab}T^{ab}$. Adopting a linear functional form $f(R,T^2) = R + \gamma T^2$, we demonstrate the existence of a wide range of wormhole solutions that satisfy all of the energy conditions without requiring fine-tuning of the model parameters. Due to the inherent complexity of the field equations, these solutions are constructed through an analytical recursive method. However, they lack a natural localization, requiring a junction with an external vacuum region. To address this, we derive the corresponding junction conditions and establish that the matching must always be smooth, precluding the formation of thin shells at the interface. Using these conditions, we match the interior wormhole geometry to an exterior Schwarzschild solution, yielding localized, static, and spherically symmetric wormholes that satisfy all the energy conditions throughout the entire spacetime. Finally, we extend our analysis to more intricate dependencies on $T^2$, demonstrating that the methodology remains applicable as long as no mixed terms between $R$ and $T^2$ are introduced.
\end{abstract}
\bodymatter
\section{Introduction}

Wormholes are theoretical structures that connect two separate spacetime manifolds or distinct regions within the same manifold. These objects have been extensively studied in the context of General Relativity (GR) \cite{morris1,Morris:1988tu,visser1,lemos1,Visser:2003yf,Kar:1995ss}, where it is found that traversable wormholes must satisfy the flaring-out condition \cite{morris1}. Combining the requirement with the Einstein field equations results in the violation of the null energy condition (NEC) and, consequently, all other energy conditions \cite{visser1,Hawking:1973uf,Sajadi:2016hko}. Matter violating the NEC, termed as \textit{exotic} matter, has limited physical relevance since it lacks experimental evidence for its existence. 

One way to avoid the use of exotic matter to sustain wormhole geometries is to examine them within the framework of modified gravity theories \cite{Sharif:2021ptz,ZeeshanGul:2023ysx,Rosa:2023guo,agnese1,nandi1,camera1,lobo1,garattini1,lobo2,garattini2,lobo3,MontelongoGarcia:2011ag,garattini3,myrzakulov1,lobo4}. In these theories, additional curvature terms contribute to sustain the geometry of the wormhole throat, ensuring its traversability, while allowing the matter content to remain non-exotic. This result can be achieved across various extensions of GR, ranging from $f\left(R\right)$ gravity and its generalizations \cite{lobo5,capozziello1,rosa1,rosa2,rosalol,rosalol2,kull1}, to models involving curvature-matter couplings \cite{garcia1,garcia2}, theories incorporating additional fundamental fields \cite{harko1,anchordoqui1}, Einstein--Cartan gravity \cite{DiGrezia:2017daq}, Gauss--Bonnet gravity \cite{bhawal1,dotti1,mehdizadeh1}, and brane-world scenarios \cite{bronnikov1,lobo6}. Regarding $f\left(R,T^2\right)$ gravity, specific solutions were first obtained using a Noether symmetry approach. However, these solutions lacked physical significance as they violated the NEC \cite{Sharif:2021ptz,ZeeshanGul:2023ysx}. 

More recently, physically viable traversable wormhole spacetimes that satisfy all energy conditions have been found \cite{Rosa:2023guo}, and these are the solutions presented in this manuscript.
More specifically, we analyze traversable wormholes in $f(R,T^2)$ gravity with $f(R,T^2) = R + \gamma T^2$. A broad class of solutions satisfying all energy conditions is found via an analytical recursive method. To ensure localization, we derive junction conditions, proving smooth matching to an external Schwarzschild solution. The approach extends to more general $T^2$ dependencies, provided no mixed terms with $R$ appear.


\section{Theory and framework}\label{sec:theory}

\subsection{Action and field equation of $f\left(R,T^2\right)$}\label{subsec:action}

In this work, we investigate wormholes within the framework of the $f(R,T^2)$ theory of gravity. The action $S$ governing this theory is given by
\begin{equation}\label{geo_action}
S=\frac{1}{2\kappa^2}\int_\Omega \sqrt{-g}f\left(R,T^2\right)d^4x+\int_\Omega \sqrt{-g}\mathcal{L}_m d^4x,
\end{equation}
where $\kappa^2=8\pi $, where we adopt a geometrized unit system with $c=G=1$. Here, $\Omega$ represents a spacetime manifold, and $g$ is the determinant of the metric tensor $g_{ab}$. The function $f(R,T^2)$ is an arbitrary function of the Ricci scalar $R=g^{ab}R_{ab}$, where $R_{ab}$ is the Ricci tensor, and the scalar $T^2=T_{ab}T^{ab}$, where $T_{ab}$ denotes the energy-momentum tensor. The term $\mathcal{L}_m$ corresponds to the matter Lagrangian, and the energy-momentum tensor is defined as usual by
\begin{equation}\label{def_tab}
T_{ab}=-\frac{2}{\sqrt{-g}}\frac{\delta\left(\sqrt{-g}\mathcal{L}_m\right)}{\delta g^{ab}}.
\end{equation}

Varying the action~\eqref{geo_action} with respect to the metric $g_{ab}$, we obtain the modified gravitational field equations
\begin{equation}\label{geo_field}
f_R R_{ab}-\frac{1}{2}g_{ab}f-\left(\nabla_a\nabla_b-g_{ab}\Box\right)f_R=8\pi T_{ab}-f_{T^2} \Theta_{ab},
\end{equation}
where we use the shorthand notation $f_R \equiv \partial f / \partial R$ and $f_{T^2}=\partial f/\partial T^2$. Here, $\nabla_a$ denotes the covariant derivative, and $\Box=g^{ab}\nabla_a\nabla_b$ is the d'Alembert operator, both defined with respect to the metric $g_{ab}$. The term $\Theta_{ab}$ arises from the variation of $T_{ab}$ given by
\begin{equation}
\Theta_{ab} = \frac{\delta T^2}{\delta g^{ab}}.
\end{equation}
Choosing a specific matter Lagrangian $\mathcal{L}_m$ or, equivalently, a particular form of the energy-momentum tensor $T_{ab}$, the explicit expression for the auxiliary tensor $\Theta_{ab}$ is determined. Taking the covariant derivative of Eq. \eqref{geo_field} then yields the conservation equation 
\begin{equation}\label{geo_conservation}
8\pi \nabla_bT^{ab}=\nabla_b\left(f_{T^2} \Theta^{ab}\right)+f_R\nabla_bR^{ab}-\frac{1}{2}g^{ab}\nabla_bf.
\end{equation}  
This equation reveals that, in this theory, the energy-momentum tensor $T_{ab}$ is not necessarily conserved, meaning that in general, $\nabla_b T^{ab} \neq 0$. This feature distinguishes it from GR.

In this work, we assume that the function $f\left(R,T^2 \right)$ is both separable and linear in terms of $R$ and $T^2$, that is
\begin{equation}
    f\left(R,T^2\right) = R + \gamma T^2,
\end{equation}  
where $\gamma$ represents a coupling constant. Thus, the field equations \eqref{geo_field} and the conservation equation \eqref{geo_conservation} take the following form:
\begin{equation}\label{field1}
G_{ab} = 8\pi T_{ab} - \gamma\left(\Theta_{ab} - \frac{1}{2} g_{ab} T^2 \right),
\end{equation}  
\begin{equation}\label{conservation1}
8\pi \nabla_b T^{ab} = \gamma \nabla_b \left(\Theta^{ab} - \frac{1}{2} g^{ab} T^2 \right),
\end{equation}  
where we have introduced the Einstein tensor $G_{ab} = R_{ab} - \frac{1}{2} R g_{ab}$. Possible generalizations of these assumptions are discussed later in this manuscript.

\subsection{Wormhole geometry and matter distribution}

We begin by considering a static, spherically symmetric traversable wormhole metric in spherical coordinates $\left(t, r, \theta, \varphi\right)$, expressed as
\begin{equation}\label{def_metric}
ds^2 = -e^{\zeta(r)}dt^2 + \left(1 - \frac{b(r)}{r}\right)^{-1}dr^2 + r^2 \left( d\theta^2 + \sin^2\theta d\varphi^2 \right),
\end{equation}
where $\zeta(r)$ is the redshift function, $b(r)$ is the shape function. To ensure that the wormhole is traversable, the redshift function must be finite throughout the entire spacetime, i.e., $|\zeta(r)| < \infty$, to prevent event horizons. Additionally, the flaring-out condition at the wormhole throat $r=r_0$ is required \footnote{The flaring-out condition close to the throat takes the form of $(b - b'r)/b^2 > 0$ \cite{morris1}.}, resulting in the following two boundary conditions: $b(r_0)=r_0$, and $b'(r_0)<1$.

Upon these requirements, we consider two general families of solutions for the functions $\zeta(r)$ and $b(r)$, given by
\begin{equation}\label{zbfunctions}
\zeta(r) = \zeta_0 \left(\frac{r_0}{r}\right)^\alpha, \qquad b(r) = r_0 \left(\frac{r_0}{r}\right)^\beta,
\end{equation}
where $\zeta_0$ is an arbitrary constant to be specified, and $\alpha$ and $\beta$ are arbitrary positive exponents. 

For the matter sector, we assume that the matter distribution is described by an anisotropic perfect fluid. Therefore, the energy-momentum tensor $T_{ab}$ takes the form
\begin{equation}\label{def_matter}
T_a^b=\text{diag}\left(-\rho,p_r,p_t,p_t\right),
\end{equation}
where $\rho \equiv \rho(r)$ is the energy density, $p_r \equiv p_r(r)$ represents the radial pressure, and $p_t \equiv p_t(r)$ denotes the tangential pressure. 

Based on these assumptions, and with the choice in Eq.\eqref{zbfunctions}, the field quation \eqref{field1} leads to three independent components, which are given by the following expressions
\begin{eqnarray}\label{eqrho}
8\pi\rho &=& \frac{\gamma}{6}\left(p_r^2-2p_t^2-3\rho^2-8p_rp_t-8p_r\rho-16p_t\rho\right) - \frac{\beta}{r^2}\left(\frac{r_0}{r}\right)^{\beta+1},
\end{eqnarray}
\begin{eqnarray}\label{eqpr}
8\pi p_r &=& \frac{\gamma}{6}\left(p_r^2+2p_t^2-3\rho^2-12p_rp_t+4p_r\rho-4p_t\rho\right)
	\nonumber \\
&&  -\frac{1}{r^2}\left(\frac{r_0}{r}\right)^{\beta+1}-\frac{\alpha\zeta_0}{r^2}\left(\frac{r_0}{r}\right)^\alpha\left[1-\left(\frac{r_0}{r}\right)^{\beta+1}\right],
\end{eqnarray}
\begin{eqnarray}\label{eqpt}
8\pi p_t &=&-\frac{\gamma}{6}\left(p_r^2+6p_t^2+3\rho^2+2p_rp_t+2p_r\rho-2p_t\rho\right)
	\nonumber \\
&& +\frac{1+\beta}{2r^2}\left(\frac{r_0}{r}\right)^{\beta+1}
+\frac{\alpha^2\zeta_0^2}{4r^2}\left(\frac{r_0}{r}\right)^{2\alpha}\left[1-\left(\frac{r_0}{r}\right)^{\beta+1}\right]
\nonumber \\
&&	+\frac{\alpha\zeta_0}{4r^2}\left(\frac{r_0}{r}\right)^\alpha\left[2\alpha-\left(1+2\alpha+\beta\right)\left(\frac{r_0}{r}\right)^{\beta+1}\right].
\end{eqnarray}
The Eqs. \eqref{eqrho}--\eqref{eqpt} constitute a system of three equations with three unknowns: $\rho$, $p_r$, and $p_t$. Each of these equations is quadratic in its respective unknown, which suggests that the system could have up to eight independent solutions. However, the nature of these solutions may vary, with some solutions potentially being complex, depending on the particular values of the involved parameters.

\subsection{Wormhole solutions}\label{subsec:whsolutions}

Due to the complexity of the system of Eqs.\eqref{eqrho}--\eqref{eqpt}, obtaining explicit analytical solutions for $\rho$, $p_r$, and $p_t$ is not feasible. However, using a recursive approach, we can determine these quantities by starting with specific values for the free parameters and solving the system at the initial radius $r_0$. From there, we incrementally calculate $\rho$, $p_r$, and $p_t$ at larger radii, up to a sufficient large radius $r$. This method also works for any $f\left(R, T^2\right)$ function with higher powers of $T^2$ and no cross terms of $R T^2$, although the set of solutions is larger. 

Among the solutions obtained, we focus on those that are astrophysically relevant, meaning their matter components satisfy the energy conditions. For a diagonal energy-momentum tensor $T_{ab}$ as described in Eq. \eqref{def_matter}, these energy conditions are expressed as follows
\begin{equation}
    \rho + p_r \geq 0, \quad \rho + p_t \geq 0, \quad \rho \geq 0, \quad \rho + p_r + 2 p_t \geq 0, \quad \rho \geq |p_r|, \quad \rho \geq |p_t|.
\end{equation}
From the set of eight solutions for the matter quantities, any solutions that violate any of these energy conditions are excluded, and only those that satisfy all the energy conditions are studied.

\begin{figure}[htbp]
  \centering
  \includegraphics[width=0.48\textwidth]{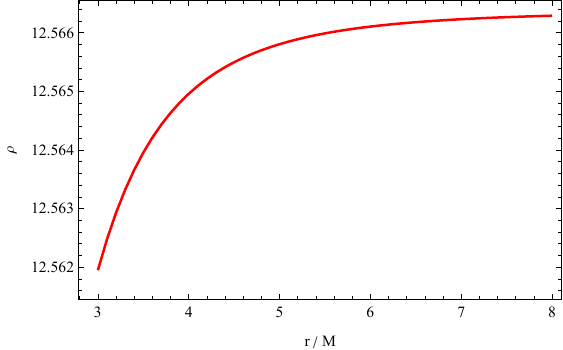}
 \includegraphics[width=0.48\textwidth]{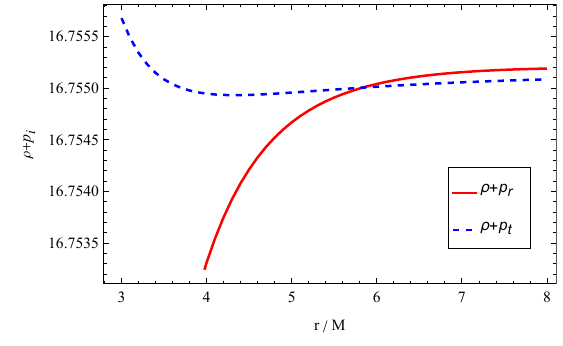}\\
  \includegraphics[width=0.48\textwidth]{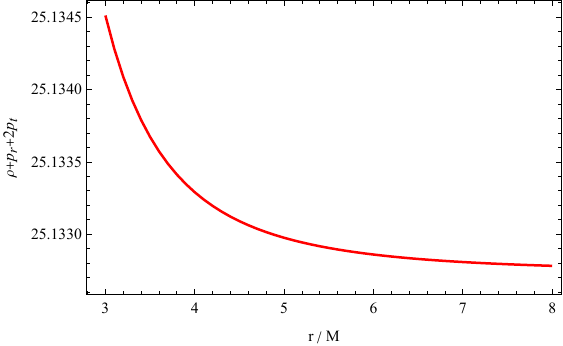}
  \includegraphics[width=0.48\textwidth]{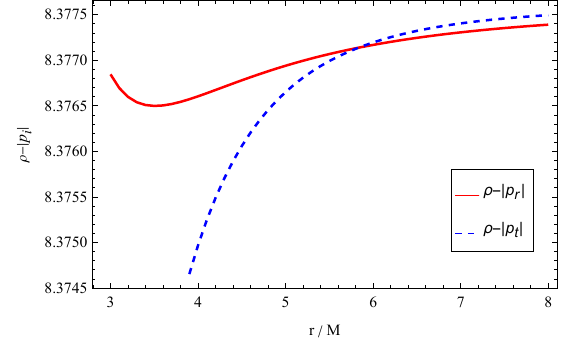}
  \caption{Matter density $\rho$, along with the expressions $\rho + p_r$, $\rho + p_t$, $\rho + p_r + 2p_t$, $\rho - |p_r|$, and $\rho - |p_t|$, are shown for the parameter set $\alpha = \beta = -\gamma = 1$, $r_0 = 3M$, and $\zeta_0 = -\frac{6}{5}$.}
  \label{fig:solution}
\end{figure}

As a specific example, we consider the parameter combination $\alpha = \beta = -\gamma = 1$, $r_0 = 3M$, and $\zeta_0 = -\frac{6}{5}$.\footnote{
While the choice of $\zeta_0$ is somewhat arbitrary at this stage, this particular value is selected for reasons explained in the following section. Other values of $\zeta_0$, including positive ones, would lead to qualitatively similar solutions.} The matter density $\rho$, along with the combinations $\rho + p_r$, $\rho + p_t$, $\rho + p_r + 2p_t$, $\rho - |p_r|$, and $\rho - |p_t|$, are plotted in Figure \ref{fig:solution} for the solution that satisfies all the energy conditions discussed above.

As in linear $f(R,T)$ gravity \cite{kull1}, our analysis of solutions satisfying all the energy conditions in linear $f(R,T^2)$ confirms that they only exist for negative values of $\gamma$. However, for higher-order terms, $f(R,T^2) = R + \gamma T^2 + \sigma (T^2)^n$, such solutions can exist for positive values of $\gamma$, provided that $\sigma$ remains negative. In both cases, the matter components remain nonzero across all $r$, despite asymptotic flatness, as seen in Figure \ref{fig:solution}, thus necessitating a matching with an exterior vacuum spacetime at a finite radius. This issue is addressed in the following section.

\section{Junction Conditions and Matching}\label{sec:matching}

To obtain physically meaningful spacetime solutions describing localized objects, junction conditions must be applied to match the interior and exterior spacetimes at a finite radius. First derived in GR \cite{israel1}, these conditions have been widely used in studies of traversable wormholes \cite{visser2,visser3,Lobo:2004rp,Lobo:2004id,Lobo:2005us,Lobo:2005yv}, fluid stars \cite{schwarzschild1,rosafluid,rosafluid2}, and gravitational collapse \cite{oppenheimer1,rosa112}. Since these conditions depend on the underlying theory, various works have explored them in modified gravity (see \cite{Rosa:2023tph} for a review), including $f(R)$ gravity \cite{senovilla1,Vignolo:2018eco,Reina:2015gxa,Deruelle:2007pt,Olmo:2020fri,rosafrt,rosafrt2}, theories with additional fundamental fields \cite{suffern,Barrabes:1997kk,Padilla:2012ze,Casado-Turrion:2023omz}, and metric-affine gravity \cite{delaCruz-Dombriz:2014zaa,Arkuszewski:1975fz,amacias}. More recently, junction conditions have been studied in $f(R,T^2)$ gravity \cite{Rosa:2023guo}, which we focus on here by deriving the conditions and performing the matching between the interior wormhole with an exterior vacuum solution.

\subsection{Junction conditions of linear $f\left(R,T^2\right)$ gravity}\label{subsec:jc}

We first derive the junction conditions for the linear case, $f(R,T^2) = R + \gamma T^2$, then briefly extend the analysis to higher-order terms, $f(R,T^2) = R + \gamma T^2 + \sigma \left(T^2\right)^n$.

Let us consider a spacetime manifold $\Omega$ divided into two regions $\Omega^\pm$, each with metric tensors $g_{ab}^\pm$ in coordinates $x^a_\pm$. The interior and exterior spacetimes are $\Omega^-$ and $\Omega^+$, respectively, separated by a three-dimensional hypersurface $\Sigma$ with induced metric $h_{\alpha \beta}$, expressed in coordinates $y^\alpha$, where greek indices exclude the directional normal to $\Sigma$. The projection tensors mapping $\Omega$ onto $\Sigma$ are $e^a_\alpha=\partial x^a/\partial y^\alpha$. The normal vector to $\Sigma$ is $n_a=\epsilon \partial_a l$, where $l$ is an affine parameter along geodesics normal to $\Sigma$, and $\epsilon = 1, -1, 0$ for space-like, time-like, and null geodesic congruences, respectively. By construction, the condition $n^a e_a^\alpha=0$ is always satisfied. With this notation, the induced metric $h_{\alpha \beta}$ and extrinsic curvature $K_{\alpha \beta}$ of $\Sigma$ are given by  
\begin{equation}\label{defhK}
h_{\alpha\beta}=e^a_\alpha e^b_\beta g_{ab}, \qquad K_{\alpha\beta}=e^a_\alpha e^b_\beta\nabla_a n_b.
\end{equation}
The junction conditions are obtained using the distribution formalism, where any quantity $X$ and its derivative $\nabla_a X$ are expressed through distribution functions as  
\begin{equation}\label{distX}
X=X^+\theta\left(l\right)+X^-\theta\left(-l\right),
\end{equation}
\begin{equation}\label{distdX}
\nabla_a X=\nabla_a X^+\theta\left(l\right)+\nabla_a X^-\theta\left(-l\right)+\epsilon n_a\left[X\right]\delta\left(l\right),
\end{equation}
where $X^\pm$ denotes $X$ in the respective spacetimes $\Omega^\pm$; $\theta(l)$ is the Heaviside distribution function, defined as $\theta(l) = 0$ for $l < 0$, $\theta(l) = 1$ for $l > 0$, and $\theta(l) = \frac{1}{2}$ for $l = 0$; and $\delta(l) = \partial_l \theta(l)$ is the Dirac delta distribution. The discontinuity of $X$ across $\Sigma$ is denoted by $\left[X\right]$, defined as 
\begin{equation}
\left[X\right]=X^+|_\Sigma - X^-|_\Sigma.
\end{equation}
If the quantity $X$ is continuous across $\Sigma$, then $\left[X\right] = 0$. By definition, we have $\left[n^a\right]=\left[e^a_\alpha\right]=0$.

To derive the junction conditions, we reformulate the field equations \eqref{field1} within the distribution formalism. Starting with the metric tensor $g_{ab}$, it is expressed in this framework as  
\begin{equation}\label{eq:def_metric}
g_{ab}=g_{ab}^+\theta\left(l\right)+g_{ab}^-\theta\left(-l\right).
\end{equation}
From the metric expression in Eq.\eqref{eq:def_metric}, we now compute the Christoffel symbols $\Gamma^c_{ab}$ associated with the metric $g_{ab}$. This involves computing the derivatives $\partial_c g_{ab}$, which, using Eq.\eqref{distdX}, take the form  
\begin{equation}
\partial_c g_{ab}=\partial_c g_{ab}^+\theta\left(l\right)+\partial_c g_{ab}^-\theta\left(-l\right)+\epsilon n_c\left[g_{ab}\right]\delta\left(l\right).
\end{equation}
The term proportional to $\delta(l)$ poses a challenge in defining the Riemann tensor $R^a_{\ bcd}$ within the distributional formalism. Since the Riemann tensor $R^a_{\ bcd}$ involves products of Christoffel symbols $\Gamma^c_{ab}$, it results in terms proportional to $\delta^2(l)$, which are singular in this formalism. The junction conditions are introduced precisely to eliminate such singularities from the field equations. To ensure the removal of these singularities in the Riemann tensor $R^a_{\ bcd}$, we impose the continuity of the metric $g_{ab}$ across $\Sigma$, requiring  
$\left[g_{ab}\right]=0$. Given that $\left[e^a_\alpha\right]=0$, this condition can be expressed in a coordinate-independent form by projecting both indices onto $\Sigma$, leading to
\begin{equation}\label{junction1}
\left[h_{\alpha \beta}\right]=0.
\end{equation}
This equation gives the first junction condition, ensuring the induced metric on $\Sigma$ is continuous. Consequently, the derivatives of $g_{ab}$ are expressed as
\begin{equation}\label{dmetric}
\partial_c g_{ab}=\partial_c g_{ab}^+\theta\left(l\right)+\partial_c g_{ab}^-\theta\left(-l\right).
\end{equation}
With the metric derivatives, we can now compute the Christoffel symbols, followed by the Riemann tensor and its contractions, specifically the Ricci tensor $R_{ab}$ and the Ricci scalar $R$, which remain well-defined. These are given by
\begin{equation}\label{eq:dist_Rabcd}
R_{abcd}=R^+_{abcd}\theta\left(l\right)+R^-_{abcd}\theta\left(-l\right)+\bar R_{abcd}\delta\left(l\right),
\end{equation}
\begin{equation}\label{eq:dist_Rab}
R_{ab}=R^+_{ab}\theta\left(l\right)+R^-_{ab}\theta\left(-l\right)+\bar R_{ab}\delta\left(l\right),
\end{equation}
\begin{equation}\label{eq:dist_R}
R=R^+\theta\left(l\right)+R^-\theta\left(-l\right)+\bar R\delta\left(l\right),
\end{equation}
where $\bar R_{abcd}$, $\bar R_{ab}$, and $\bar R$ represent the $\delta(l)$-proportional terms, expressed in terms of geometric quantities as  
\begin{equation}\label{eq:def_barRabcd}
\bar R_{abcd}=4\left[K_{\alpha\beta}\right]e^\alpha_{[a}n_{b]}e^\beta_{[d}n_{c]},
\end{equation}
\begin{equation}\label{eq:def_barRab}
\bar R_{ab}=-\left(\epsilon\left[K_{\alpha\beta}\right]e^\alpha_a e^\beta_b+n_a n_b \left[K\right]\right),
\end{equation}
\begin{equation}\label{eq:def_barR}
\bar R=-2\epsilon\left[K\right],
\end{equation}
where $X_{[ab]} \equiv \frac{1}{2} (X_{ab} - X_{ba})$ defines index anti-symmetrization, and $K = h^{\alpha\beta} K_{\alpha\beta}$ is the trace of the extrinsic curvature.

Regarding the matter sector, it is helpful to associate the presence of a thin shell at the hypersurface $\Sigma$ with terms that are proportional to $\delta\left(l\right)$ in the gravitational sector of the modified field equations. The energy-momentum tensor is then written as
\begin{equation}\label{eq:dist_tab}
T_{ab}=T_{ab}^+\theta\left(l\right)+T_{ab}^-\theta\left(-l\right)+S_{ab}\delta\left(l\right),
\end{equation}
where $S_{ab}=S_{\alpha\beta}e^\alpha_a e^\beta_b$, with $S_{\alpha\beta}$ being the three-dimensional energy-momentum tensor of the thin shell. To derive $T^2$ in the distributional formalism, we contract $T_{ab}$ with itself using Eq.\eqref{eq:dist_tab}, resulting in
\begin{equation}\label{eq:dist_tab2}
T^2=\left(T^2\right)^+\theta\left(l\right)+\left(T^2\right)^-\theta\left(-l\right)+\bar{\left(T^2\right)}\delta\left(l\right)+\hat{\left(T^2\right)}\delta^2\left(l\right),
\end{equation}
Note that the term proportional to $\delta^2\left(l\right)$ in Eq.\eqref{eq:dist_tab2} is singular and must be eliminated. This can only be achieved by imposing that the energy-momentum tensor of the thin shell vanishes, that is,
\begin{equation}\label{junctionS}
    S_{ab}=0.
\end{equation}
When Eq.\eqref{junctionS} is satisfied, the matching is called a smooth matching. In $f\left(R,T^2 \right)$ gravity, this is the only method that preserves the regularity of the action, unlike in other gravity theories where it is a special case of a more general thin-shell matching.

Under the constraint from Eq.\eqref{junctionS}, we project the field equations for ${f\left(R,T^2\right) = R + \gamma T^2}$ onto the hypersurface $\Sigma$ using $e^a_\alpha e^b_\beta$, leading to $\left[K_{\alpha\beta}\right]-\left[K\right]h_{\alpha\beta}=0$. Taking the trace of this equation with $h^{\alpha \beta}$ gives $\left[K\right]=0$, which, when substituted back into the original equation, results in
\begin{equation}
\left[K_{\alpha\beta}\right]=0.
\end{equation}
Thus, the second junction condition requires that the extrinsic curvature $K_{\alpha \beta}$ remains continuous across $\Sigma$.

In conclusion, the matching in linear $f\left(R,T^2\right)$ gravity is always smooth, meaning there is no thin shell involved. The junction conditions are the same as in GR: both the induced metric $h_{\alpha \beta}$ and the extrinsic curvature $K_{\alpha \beta}$ must be continuous across $\Sigma$, i.e.,
\begin{equation}\label{junction}
\left[h_{\alpha \beta}\right]=0, \qquad \left[K_{\alpha \beta}\right]=0.
\end{equation}

When considering higher-order powers of $T^2$ in the function, the field equations involve terms like $\left(T^2\right)^{n-1}\Theta_{ab}$ and $\left(T^2\right)^n$. In the linear case, we showed that for $T^2$ to be well-defined in the distributional formalism, the matching must be smooth, implying $S_{ab} = 0$. In fact, this condition ensures that both $T^2$ and the auxiliary tensor $\Theta_{ab}$ remain regular, with terms proportional to $\theta(l)$ and excluding $\delta(l)$. Consequently, the products of $T^2$ and $\Theta_{ab}$, as well as the powers of $\left(T^2\right)^n$, preserve this regularity. As a result, higher powers of $T^2$ do not introduce additional junction conditions, provided that mixed $RT^2$ terms are absent.

\subsection{Matching with an exterior vacuum spacetime}

We now apply the junction conditions to match the interior wormhole spacetime with an exterior spherically symmetric and static vacuum solution. The metric expressions for the interior and exterior spacetimes are
\begin{equation}\label{metrici}
ds_-^2=-C e^{\zeta_0\left(\frac{r_0}{r}\right)^\alpha}dt^2+\left[1-\left(\frac{r_0}{r}\right)^{\beta+1}\right]^{-1}dr^2+r^2d\Omega^2,
\end{equation}
\begin{equation}\label{metrice}
ds_+^2=-\left(1-\frac{2M}{r}\right)dt^2+\left(1-\frac{2M}{r}\right)^{-1}dr^2+r^2d\Omega^2,
\end{equation}
respectively. The metric in Eq.\eqref{metrici} is derived from Eq.\eqref{def_metric} using the ansatz for the redshift and shape functions in Eq.\eqref{zbfunctions}. The constant $C$ is introduced for convenience to ensure that the time coordinates of the interior and exterior metrics are continuous. Meanwhile, the metric in Eq.\eqref{metrice} corresponds to the Schwarzschild solution with mass $M$ \cite{Schwarzschild:1916uq}.

The analysis is simplified by first considering the second junction condition \eqref{junction}. Due to the spherical symmetry of the metrics under consideration, the extrinsic curvatures $K_{\alpha \beta}^\pm$ have two independent components: $K_{00}$ and $K_{\theta\theta} = K_{\phi\phi} \sin^2\theta$. This leads to the constraints $\left[K_{00}\right] = 0$ and $\left[K_{\theta\theta}\right] = 0$, expressed as
\begin{equation}\label{cond1}
\frac{\alpha\zeta_0}{2}\left(\frac{r_0}{r}\right)^\alpha\sqrt{1-\left(\frac{r_0}{r}\right)^{\beta+1}}+\frac{M}{r}\sqrt{\frac{r}{r-2M}}=0,
\end{equation}
\begin{equation}\label{cond2}
\sqrt{1-\left(\frac{r_0}{r}\right)^{\beta+1}}=\sqrt{1-\frac{2M}{r}},
\end{equation}
respectively. Solving the second equation for the radius $r$ using Eq.\eqref{cond2}, we obtain unique real solutions for $M > 0$ and $r_0 > 0$, given by
\begin{equation}\label{rsigma}
r_\Sigma=\left(2M\right)^{-\frac{1}{\beta}}\left(r_0\right)^{1+\frac{1}{\beta}},
\end{equation}
This determines the matching radius $r_\Sigma$, which must satisfy $r_\Sigma > 2M$ to avoid event horizons. This implies $r_0 > 2M$ for any $\beta \geq 1$. Substituting the solutions for $r_\Sigma$ into Eq.\eqref{cond1}, we solve for $\zeta_0$, resulting in the expression
\begin{equation}\label{zsigma}
\zeta_0=\frac{\left(2M\right)^{\frac{1-\alpha+\beta}{\beta}}\left(r_0\right)^{\frac{\alpha}{\beta}}}{\alpha\left[\left(2M\right)^{1+\frac{1}{\beta}}-\left(r_0\right)^{1+\frac{1}{\beta}}\right]}.
\end{equation}

Since $r_0 > 2M$, for $\alpha \geq 1$ and $\beta \geq 1$, the condition $\zeta_0 < 0$ holds. This result is expected, as negative values of $\zeta_0$ guarantee that the derivative of $g_{00}$ has consistent signs in both the interior and exterior metrics, which is essential for achieving a smooth matching.

Next, we examine the first junction condition stated in Eq.\eqref{junction}. Given that the angular components of the metrics in Eqs.\eqref{metrici} and \eqref{metrice} are identical, the angular parts of the induced metric $h_{\alpha\beta}$ are straightforwardly continuous. Consequently, the continuity condition $\left[h_{00}\right]=0$ can be analyzed separately and is expressed as follows
\begin{equation}\label{cond3}
C e^{\zeta_0\left(\frac{r_0}{r}\right)^\alpha}=\left(1-\frac{2M}{r}\right).
\end{equation}
Using the results from the second junction condition, specifically the radius $r_\Sigma$ from Eq.\eqref{rsigma} and the corresponding $\zeta_0$ from Eq.\eqref{zsigma}, we substitute these into Eq.\eqref{cond3} to solve for the constant $C$, obtaining
\begin{equation}\label{csigma}
C=\left[1-\left(\frac{2M}{r_0}\right)^{1+\frac{1}{\beta}}\right]e^{-\alpha\left[\left(\frac{r_0}{2M}\right)^{1+\frac{1}{\beta}}-1\right]}.
\end{equation}
Since $r_0 > 2M$, the constant $C$ remains positive for all $\alpha \geq 1$ and $\beta \geq 1$, ensuring the correct metric signature.  

In summary, for $r_0 > 2M$, $\alpha \geq 1$, and $\beta \geq 1$, the second junction condition $\left[K_{\alpha\beta}\right] = 0$ determines $r_\Sigma$ (Eq.\eqref{rsigma}) and $\zeta_0$ (Eq.\eqref{zsigma}), while the first condition ${\left[h_{\alpha\beta}\right] = 0}$ fixes $C$ (Eq.\eqref{csigma}), ensuring the continuity of the full spacetime metric.

\begin{figure}[htbp]
  \centering
  \includegraphics[width=0.48\textwidth]{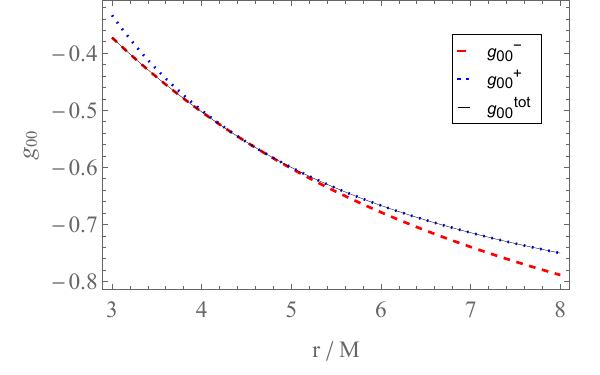}
  \includegraphics[width=0.48\textwidth]{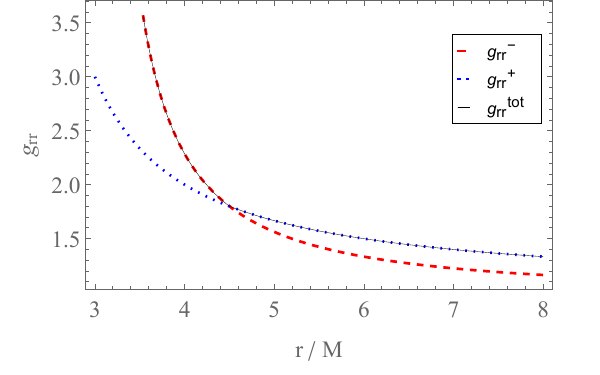}\\
  \caption{The interior wormhole spacetime from Eq.\eqref{metrici} (red dashed curve) and the exterior Schwarzschild spacetime from Eq.\eqref{metrice} (blue dotted curve) are shown for $\beta = 1$, $r_0 = 3M$, and $\alpha = 1$. The thin black line represents the matched solution at $r = r_\Sigma$, illustrating $g_{00}^{\text{tot}}$ (left panel) and $g_{rr}^{\text{tot}}$ (right panel).}
    \label{fig:matching}
\end{figure}

As an example, let us consider $r_0 = 3M$ and $\alpha = \beta = 1$, for which Eq.\eqref{rsigma} gives $r_\Sigma = \frac{9}{2}M$, Eq.\eqref{zsigma} yields $\zeta_0 = -\frac{6}{5}$, and Eq.\eqref{csigma} gives $C = \frac{5}{9} e^{\frac{4}{5}}$. The left panel of Figure \ref{fig:matching} shows the $g_{00}$ components of the interior, exterior, and matched metrics, illustrating the smooth transition of $g_{00}$, thus ensuring the continuity of both the induced metric and extrinsic curvature.

Examining the radial metric component $g_{rr}$, right panel of Figure \ref{fig:matching}, we observe that although $g_{rr}$ is not explicitly constrained by the junction conditions, since both $h_{\alpha \beta}$ and $K_{\alpha \beta}$ are three-dimensional tensors on $\Sigma$, it remains continuous but not differentiable at $r = r_\Sigma$. This continuity of $g_{rr}$ is expected due to its dependence on the mass function within a spherical hypersurface of radius $r$, given by ${g_{rr} = \left( 1 - \frac{2m(r)}{r} \right)^{-1}}$, from which we obtain $m(r) = \frac{r_0}{2} \left( \frac{r_0}{r} \right)^\beta$ (see Eqs. \eqref{zbfunctions} and \eqref{metrici}). Since the transition between the interior and exterior spacetimes is smooth, the mass function $m(r)$ remains continuous at $r_\Sigma$, therefore ensuring the continuity of the $g_{rr}$ component of the metric.

\section{Conclusion}

In conclusion, this analysis of traversable wormholes in $f(R,T^2)$ gravity with a linear dependence on $R$ and $T^2$ reveals numerous solutions satisfying all energy conditions, emphasizing their physical relevance. While the spacetime metrics of these solutions are asymptotically flat, the theory allows for non-localized matter distributions, which can be localized via junction conditions. These conditions, derived here, permit only a smooth matching, since the scalar $T^2$ becomes singular in the presence of a thin shell. As a result, they reduce to the junction conditions of GR, requiring continuity of the induced metric and extrinsic curvature at the hypersurface separating the interior and exterior spacetime regions. The matching procedure then yields localized wormhole solutions satisfying all energy conditions, enhancing their astrophysical significance. Moreover, the methods presented can be extended to more general $f(R,T^2)$ models, provided no crossed terms between $R$ and $T^2$ are present. Furthermore, the smooth matching requirement ensures that, in the absence of crossed terms, no additional junction conditions arise, allowing for the effective localization of solutions as in the linear $T^2$ case.

\section{Acknowledgments}
JLR acknowledges the European Regional Development Fund and the programme Mobilitas Pluss for
financial support through Project No.~MOBJD647, and project No.~2021/43/P/ST2/02141 co-funded by the Polish National Science Centre and the European Union Framework Programme for Research and Innovation Horizon 2020 under the Marie Sklodowska-Curie grant agreement No. 94533.
N.G. and F.S.N.L. acknowledge funding from the Funda\c{c}\~{a}o para a Ci\^{e}ncia e a Tecnologia (FCT) research grants UIDB/04434/2020 and UIDP/04434/2020.
F.S.N.L. acknowledges support from the FCT Scientific Employment Stimulus contract with reference CEECINST/00032/2018, and funding from the FCT research grant PTDC/FIS-AST/0054/2021.


\end{document}